\documentclass[12pt]{article}
\usepackage{epsfig}
\begin{document}
\title{On heavy Majorana neutrinos as a source of the
highest energy cosmic rays}
\author{D. PALLE \\
Zavod za teorijsku fiziku \\
Institut Rugjer Bo\v skovi\'c \\
Po\v st. Pret. 180, HR-10002 Zagreb, CROATIA}
\date{ }
\maketitle

\vspace{5 mm}
{\it 
Cosmic ray events beyond the Greisen-Zatsepin-Kuzmin cut-off
represent a great challenge for particle physics and
cosmology.
We show that the physics of heavy Majorana neutrinos,
well defined by their masses, cross sections and lifetimes,
could explain the highest energy cosmic rays as a consequence
of the galactic annihilation of heavy neutrinos as 
cold dark matter particles.
Galactic nuclei accelerators,
colliding neutron stars (black holes) or shocks from the collapsed objects
could produce ultra high energy cosmic rays as heavy
neutrinos beyond the mass threshold at an arbitrary
cosmic distance.
We comment and also analyse the DAMA results with regard to heavy neutrinos
as galactic halo CDM particles.
}
\vspace{5 mm}

It seems that cosmic ray
 events accumulated at Yakutsk, Haverah Park, Fly's Eye,
 HiRes and AGASA beyond the Greisen-Zatsepin-Kuzmin (GZK) cut-off
\cite{Greisen} are definitely established as  
an observational fact, but also as a theoretical problem.
The gyroradius of the particle in the galactic magnetic
field and available galactic energy sources are in conflict
with these ultra high energy (UHE) events, which suggests its extragalactic
origin.
However, the GZK absorption length $\lambda_{\gamma p}
\simeq 10Mpc$ cannot be reconciled with protons as 
extragalactic UHE cosmic rays. 

From a variety of attempts to solve the problem, we mention
only two that we find interesting and plausible: light
neutrinos \cite{Weiler} and superheavy particles \cite{Berezinsky}.
A detailed analysis of the Z-burst scenario \cite{Kalashev}
(interaction of UHE and relic light neutrinos)
shows that it is difficult to find an astrophysical mechanism
that could simultaneously
explain neutrino, nucleon and photon fluxes 
because they are subject of severe
constraints.
The proposed existence of a superheavy particle
(which does not belong to any well-founded particle model or
theory)
and the attempt to explain UHE data through
the decay or the annihilation of the superheavy particle
requires rather unusual cross sections and a large
flux of photons
\cite{Berezinsky,Blasi}.

In this paper we discuss the role of heavy 
Majorana neutrinos in order to explain UHE
cosmic rays.   
Let us recall the main features of heavy
neutrinos as a key ingredient of the theory
described in Ref. \cite{Palle1}.
We construct a theory free of the SU(2)
global anomaly and free of the ultraviolet singularity
in order to
render this theory mathematically consistent.
Light and heavy Majorana neutrinos are new key ingredients
of the particle spectrum
that play essential roles in cosmology \cite{Palle2}.
It is important to emphasize the absence of Higgs scalars,
as an essential fact to have cosmologically stable
heavy neutrinos \cite{Palle2}.
The Nambu-Goldstone scalars and heavy neutrinos are in
a strong coupling regime.

Let us now focus on UHE cosmic ray events.
Our first observation relies on the energies of
UHE events. The laboratory energy of UHE events 
corresponds to the following centre of mass energy: 
$E_{CM}\simeq\sqrt{2E_{LAB}m_{p}}=750 TeV$ for $E_{LAB}=3\times 10^{20}eV$.
One can easily imagine astrophysical processes such as
galaxy nuclei accelerators, shocks of collapsed stellar
objects or colliding neutron stars \cite{Biermann}, 
so energetic as to produce
heavy neutrinos $N_{\mu}$ or $N_{\tau}$ with 
masses $m_{N_{i}}={\cal O}(100 TeV)$ and lifetimes $\tau_{N_{i}}=
{\cal O}(10^{25}s)$ \cite{Palle2}.
They can propagate through very large cosmological
distances because they are neutral and cosmologically
stable ($\tau_{N_{i}} > 10^{17}s$) with no energy loss and
avoiding the GZK cut-off.
These cosmic UHE messsengers can, at the end of its 
cosmic voyage, hit the relic cold dark matter (CDM)
particles (again heavy Majorana neutrinos)
that form the halo of our Milky Way galaxy.
As a result of annihilation,  
the Standard Model (SM) particles,
such as hadrons, light neutrinos and photons,  
besides heavy neutrinos, are produced.
 
We show in Ref. \cite{Palle2} that the Griest-Kamionkowski (GK)
unitarity bound \cite{Griest} for annihilation cross sections
could be saturated
by heavy Majorana neutrinos with mass ${\cal O}(100 TeV)$
strongly coupled to the Nambu-Goldstone scalars 
as shown in the diagram
depicted in Fig. 1.
In the strong coupling regime, the self-energy of the
Nambu-Goldstone scalar is dominantly given by the loop
in the Dyson-Schwinger (DS) equation \cite{Palle2} and the same
is also valid for the effective scalar-neutrino coupling and its 
corresponding DS equation.
We can easily find a heavy neutrino annihilation process with
scalars in the loop and two-particle final state as two
electroweak gauge bosons (see Fig. 2).
As we expect, the cross section of this process
evaluated in the 't Hooft-Feynman gauge can achieve and
even exceede the GK unitarity bound (this is due to the
approximate 
gauge dependent-calculation in the strong coupling regime)
 \cite{Palle2,Palle3}:

\begin{eqnarray*}
\sigma_{A}(N_{i}N_{i}\rightarrow ZZ;s\gg M_{Z}^{2}) = 
\frac{\tilde{\alpha}^{2}_{e}\alpha^{2}_{e}}
{2\times 64^{2}\pi\sin^{8}\theta_{W}}
\sqrt{s(s-4 m_{N}^{2})} \frac{s m_{N}^{2}}{M_{Z}^{4}}
\mid C_{0}-\bigtriangleup \Gamma_{\Lambda}\mid ^{2}, \\
m_{N}=100 TeV,\ \tilde{\alpha}_{e}=10^{-2},\ 
\sqrt{s}\simeq 300 TeV,\ \sigma_{A} \simeq 10^{-7} GeV^{-2} .
\end{eqnarray*}

Thus, although we may conclude that besides heavy neutrinos, all
other known SM particles, such as hadrons, neutrinos and
photons, are produced in the annihilation of $N_{\mu}$
or $N_{\tau}$ CDM particles, 
one cannot precisely calculate rates and distributions 
because of the strong coupling \cite{Jungman}.

From the analysis in
Ref. \cite{Blasi} we estimate the annihilation 
cross section assuming that two leading jets are produced
in the annihilation  
in the smooth component of the galactic halo:

\begin{eqnarray*}
{\cal F}&=&2\frac{d{\cal N}(E,E_{jet} = M_{X})}
{d E}< \sigma_{A}v >\int d^{3}\vec{d} \frac{n^{2}_{X}(d)}
{4\pi \mid \vec{d}-\vec{d}_{\odot} \mid ^{2}}, \\
n_{X}(d)&=&\frac{N_{0}}{d(d+d_{s})^{2}},\ N_{0} \propto M_{X}^{-1}, \\
\frac{d{\cal N}(x)}{dx}&\simeq &\frac{15}{16}x^{-3/2},\ for\ x\ll 1,\ 
x\equiv E/E_{jet}.
\end{eqnarray*}

Taking into account 
the scaling of the flux on the CDM particle mass and the fit of 
UHE cosmic ray spectra (see Fig. 1 of Ref. \cite{Blasi}),
we conclude on an order of magnitude relation
(owing to large uncertainties in $h,\Omega_{m},m_{N},v,\tilde{\alpha}_{e},
$ etc.):

\begin{eqnarray*}
{\cal F}& \propto &M_{X}^{-3/2} <\sigma_{A} v>, \\
M_{X}&=&10^{12} GeV,\ <\sigma_{A} v>=6\times 10^{-27} cm^{2} \\
\Rightarrow m_{N}&=&100 TeV,\ v=10^{-3} \Rightarrow
\sigma_{A} = 4.8\times 10^{-7} GeV^{-2} 
\end{eqnarray*}
\begin{eqnarray*}
\Rightarrow \sigma_{A}(GK\ unit.\ bound) \approx
\sigma_{A}(N_{i}N_{i}\rightarrow SM\ part.) \approx
\sigma_{A}(UHE\ cos.\ rays) .
\end{eqnarray*}

Thus we see that the physics of heavy Majorana neutrinos
can explain its present cosmic abundances, lifetimes and
UHE cosmic ray events, but also one should notice the necessity
of their existence
to understand the physics of light neutrinos \cite{Palle1}.

The other important search for the galactic CDM particle is
a direct detection of the nucleon-CDM particle scattering
for which the DAMA collaboration reported the evidence
for the annual modulation of the scattering rate
disfavouring the unmodulated behaviour by the
probability $4\times 10^{-4}$ with the NaI set-up   
\cite{Bernabei}.
The model dependent interpretation of the signal as a
spin-independent elastic nucleon-WIMP scattering 
gives for the mass $m_{WIMP}\simeq 50 GeV$, roughly,
$\sigma^{(p)}_{s}\simeq 3\times 10^{-6} pb$ \cite{Bernabei}.
Our aim is to interpret these data as a spin-independent
elastic nucleon-heavy neutrino scattering.
In Fig. 3 one can visualize a typical process that 
contributes to such an elastic scattering. However,
because of the strong coupling in the loop with the
Nambu-Goldstone scalars and lack of our knowledge
of this coupling at low-energy scales, it is impossible 
at present to
reasonably evaluate this kind of diagrams.
What we can do is the evaluation of the effective
coupling of heavy neutrinos and quarks, assuming
that the DAMA data are reliable.

The standard calculation of the total rate gives us the
folowing formula \cite{Kamionkowski}:

\begin{eqnarray*}
R = \rho_{0} \sigma^{(n)}_{s}
\phi (m_{WIMP},E_{T},A),
\end{eqnarray*}
\begin{eqnarray*}
\phi (m_{WIMP},E_{T},A)&=&[\frac{1+m_{WIMP}/m_{p}}
{1+m_{WIMP}/m_{nucl}}]^{2}
A^{2} \frac{1}{\sqrt{\pi}v_{0}m_{WIMP}m_{r}^{2}}
\int^{+\infty}_{E_{T}}dQ T(Q)F^{2}(Q), \\
m_{nucl}&=&Am_{p},\ m_{r}=\frac{m_{WIMP}m_{nucl}}
{m_{WIMP}+m_{nucl}},
\end{eqnarray*}
\begin{eqnarray*}
T(Q)&=&\frac{\sqrt{\pi}}{2}v_{0}\int^{v_{esc}}_{v_{min}}
\frac{f_{1}(v)}{v}dv, \\
F(Q)^{2}&=&[\frac{3j_{1}(qR_{1})}{qR_{1}}]^{2}
exp[-(qs)^{2}], \\
q&=&\sqrt{2m_{nucl}Q},\ R_{1}=(R_{0}^{2}-5s^{2})^{1/2},
\ R_{0}\simeq 1.2 fm A^{1/3},\ s\simeq 1 fm .
\end{eqnarray*}

Now we can compare the cross sections for different WIMP masses 
with a target consisting of two different nuclei (like
in the set-up of the DAMA experiment):

\begin{eqnarray*}
fr(A_{1}) + fr(A_{2}) = 1,\ R_{A_{i}} = 
\frac{1}{fr(A_{i})} R(total) \\
\Rightarrow R^{-1}(total) = R^{-1}_{A_{1}} + R^{-1}_{A_{1}},
\end{eqnarray*}
\begin{eqnarray*}
R(total,m_{WIMP}) = R(total,m_{N}) 
\end{eqnarray*}
\begin{eqnarray*}
\Rightarrow 
\frac{(\rho_{0}\sigma^{(n)}_{s})_{WIMP}}
{(\rho_{0}\sigma^{(n)}_{s})_{N}} = 
\frac{\phi^{-1}(m_{WIMP},E_{T},A_{1}) + 
\phi^{-1}(m_{WIMP},E_{T},A_{2})}
{\phi^{-1}(m_{N},E_{T},A_{1}) +
\phi^{-1}(m_{N},E_{T},A_{2})} .
\end{eqnarray*}

Using quark form factors as in Ref. \cite{Kamionkowski},
with the scalar cross section fit of the DAMA data, we estimate 
the effective coupling of heavy neutrino with quarks as

\begin{eqnarray*}
\sigma^{(p)}_{s}=3\times 10^{-6} pb,\ m_{WIMP}=50 GeV, \\
{\cal L}_{int}=f_{\chi q}\bar{\chi}\chi \bar{q}q,\ 
\sigma^{p}_{s}=f_{p}^{2}m_{p}^{2}=0.7^{2}f_{\chi q}^{2}m_{p}^{2}, \\
\Rightarrow f_{\chi q} = 1.33\times 10^{-7} GeV^{-2}, 
\end{eqnarray*}
\begin{eqnarray*}
{\cal L}_{int}&=&f_{Nq}\bar{N_{i}}\gamma_{5}N_{i}\bar{q}q, \\
m_{N}&=&400 TeV \Rightarrow f_{Nq}=1.09\times 10^{-5} GeV^{-2}.
\end{eqnarray*}

It appears that the effective coupling $f_{Nq}$ is close to the Fermi
constant $G_{F}=1.166\times 10^{-5} GeV^{-2}$.

Let us make some concluding remarks.
Evidently, the process like that in Fig. 3, as
an example, can
produce heavy neutrinos in very energetic astrophysical
events on even larger energy scales than those already observed
\cite{Biermann}.
Almost without any energy loss amd avoiding the GZK cut-off,
cosmologically stable ($\tau_{N_{i}} = {\cal O}(10^{25}s)$)
heavy Mayorana particles annihilate in the galactic halo
producing hadron showers, gamma rays and UHE neutrinos. 
We have shown 
the compatibility of the annihilation cross section
that fits UHE cosmic ray data with the GK unitarity bound for
heavy neutrinos of mass $m_{N_{i}} = {\cal O}(100TeV)$.
In the case of the smooth galactic halo, one should 
observe correlations of events towards higher CDM densities
at the centre of the halo. Halo models with subclumps
require even smaller cross sections \cite{Blasi}.
UHE neutrino detectors, like AMANDA (or the future ICECUBE),
have good prospects to observe UHE light neutrinos from
galactic annihilation \cite{Halzen}.
One should notice that heavy Majorana neutrinos
are not strongly self-interacting dark matter
\cite{Spergel}.

Recent measurements of SuperKamiokande and SNO confirm
the existence of lepton mixing and nonvanishing neutrino
masses. However, any theoretical explanation must also take care
of the spin-flavour flip and the magnetic field in the
solar interior to understand the solar neutrino data
\cite{Gruzinov}, as well as to account for all three neutrino flavours. 
 New data from KamLAND, MiniBOONE and K2K experiments
could help to resolve some puzzles of neutrino oscillations.
There is also an indication from the neutrinoless double beta decay
experiment that neutrinos are Majorana particles
\cite{Klapdor}.

Thus, forthcoming studies of UHE cosmic rays by Auger, EUSO and OWL
might appear as a study of heavy $N_{\mu}$ and $N_{\tau}$ 
neutrinos, while the LHC collider might discover a $N_{e}$
neutrino if $m_{N_{e}} < E_{CM}(LHC)/2=7 TeV$.
\newline

\hspace{50 mm} * * *
\newline

This work was supported by the Ministry of Science and Technology
of the Republic of Croatia under Contract No. 00980103.

\newpage

\noindent
{\bf Caption of Fig. 1}: Annihilation into heavy Majorana neutrinos;
 $\chi^{0}$ denotes the Nambu-Goldstone scalar. 
\\
{\bf Caption of Fig. 2}: Example of annihilation into SM particles;
 $\chi^{0}$ denotes the Nambu-Goldstone scalar, $V_{i}$ denotes
 the EW gauge boson. 
\\
{\bf Caption of Fig. 3}: Example of heavy neutrino-quark elastic scattering;
 $\chi^{0}$ denotes the Nambu-Goldstone scalar, $V_{i}$ denotes
 the EW gauge boson.
\newpage
\begin{figure}
\epsfig{figure=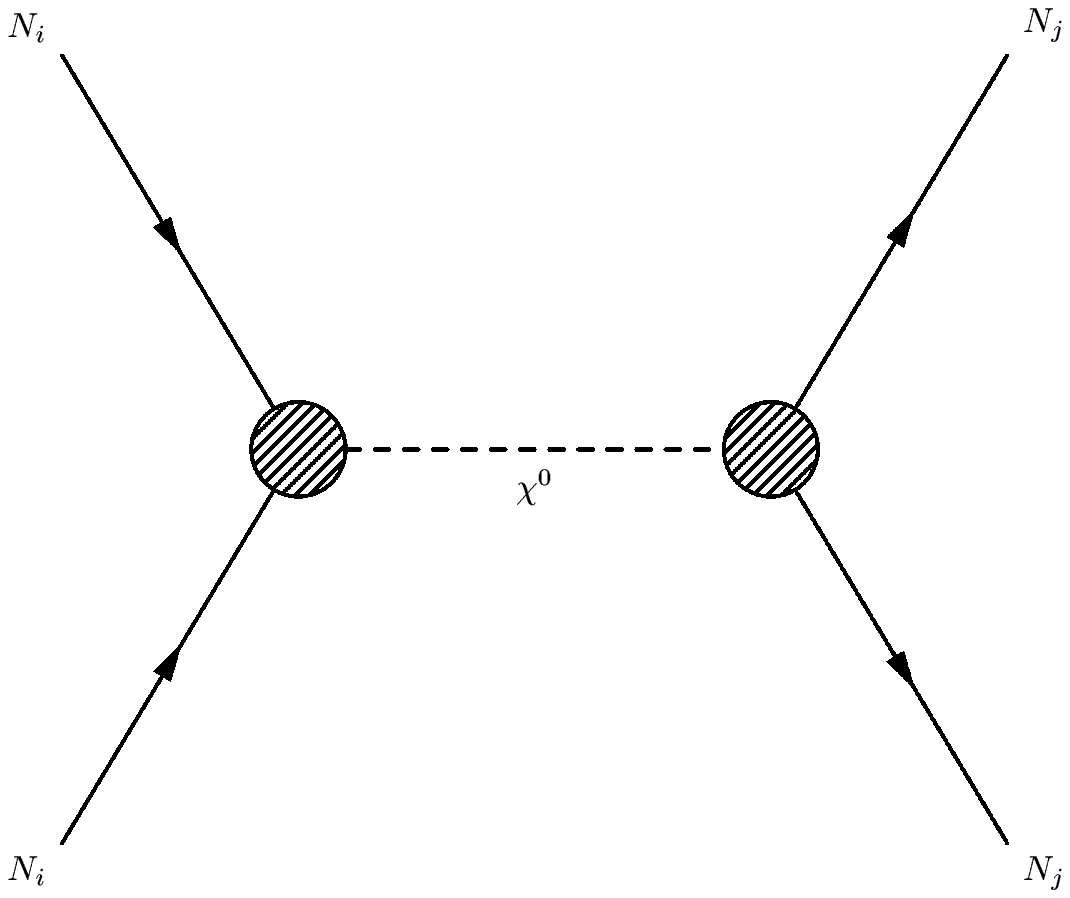, height=260 mm, width=200 mm}
\end{figure}
\newpage
\begin{figure}
\epsfig{figure=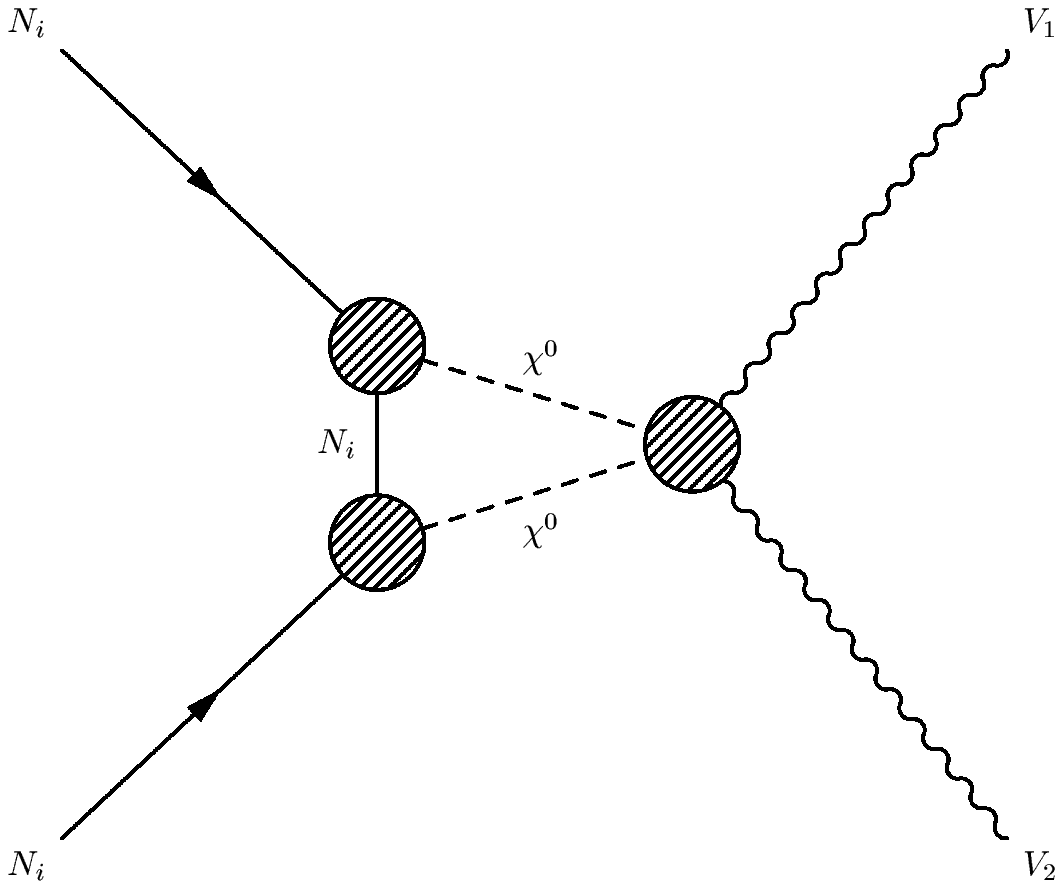, height=260 mm, width=200 mm}
\end{figure}
\newpage
\begin{figure}
\epsfig{figure=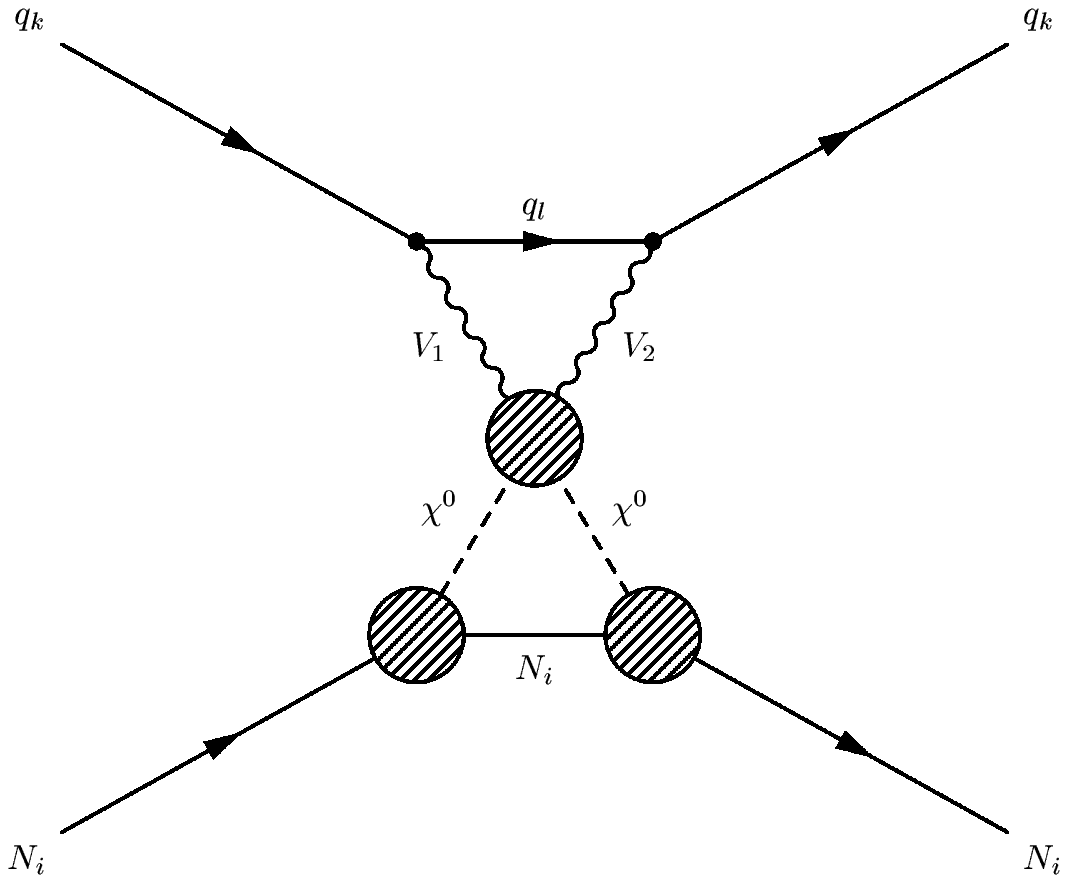, height=260 mm, width=200 mm}
\end{figure}

\end{document}